\documentstyle[epsf,12pt]{article}
\begin{document}

\catcode`@=11
\long\def\@caption#1[#2]#3{\par\addcontentsline{\csname
  ext@#1\endcsname}{#1}{\protect\numberline{\csname
  the#1\endcsname}{\ignorespaces #2}}\begingroup
    \small
    \@parboxrestore
    \@makecaption{\csname fnum@#1\endcsname}{\ignorespaces #3}\par
  \endgroup}
\catcode`@=12
\newcommand{\newc}{\newcommand}
\newc{\gsim}{\lower.7ex\hbox{$\;\stackrel{\textstyle>}{\sim}\;$}}
\newc{\lsim}{\lower.7ex\hbox{$\;\stackrel{\textstyle<}{\sim}\;$}}
\newc{\gev}{\,{\rm GeV}}
\newc{\mev}{\,{\rm MeV}}
\newc{\ev}{\,{\rm eV}}
\newc{\kev}{\,{\rm keV}}
\newc{\tev}{\,{\rm TeV}}
\newc{\mz}{m_Z}
\newc{\mpl}{M_{Pl}}
\newc{\chifc}{\chi_{{}_{\!F\!C}}}
\newc\order{{\cal O}}
\newc\CO{\order}
\newc\CL{{\cal L}}
\newc\CY{{\cal Y}}
\newc\CH{{\cal H}}
\newc\CM{{\cal M}}
\newc\CF{{\cal F}}
\newc\CD{{\cal D}}
\newc\CN{{\cal N}}
\newc{\eps}{\epsilon}
\newc{\re}{\mbox{Re}\,}
\newc{\im}{\mbox{Im}\,}
\newc{\invpb}{\,\mbox{pb}^{-1}}
\newc{\invfb}{\,\mbox{fb}^{-1}}
\newc{\yddiag}{{\bf D}}
\newc{\yddiagd}{{\bf D^\dagger}}
\newc{\yudiag}{{\bf U}}
\newc{\yudiagd}{{\bf U^\dagger}}
\newc{\yd}{{\bf Y_D}}
\newc{\ydd}{{\bf Y_D^\dagger}}
\newc{\yu}{{\bf Y_U}}
\newc{\yud}{{\bf Y_U^\dagger}}
\newc{\ckm}{{\bf V}}
\newc{\ckmd}{{\bf V^\dagger}}
\newc{\ckmz}{{\bf V^0}}
\newc{\ckmzd}{{\bf V^{0\dagger}}}
\newc{\X}{{\bf X}}
\newc{\bbbar}{B^0-\bar B^0}
\def\bra#1{\left\langle #1 \right|}
\def\ket#1{\left| #1 \right\rangle}
\newc{\sgn}{\mbox{sgn}\,}
\newc{\m}{{\bf m}}
\newc{\msusy}{M_{\rm SUSY}}
\newc{\munif}{M_{\rm unif}}
\newc{\slepton}{{\tilde\ell}}
\newc{\Slepton}{{\tilde L}}
\newc{\sneutrino}{{\tilde\nu}}
\newc{\selectron}{{\tilde e}}
\newc{\stau}{{\tilde\tau}}
%
%
\def\NPB#1#2#3{Nucl. Phys. {\bf B#1} (19#2) #3}
\def\PLB#1#2#3{Phys. Lett. {\bf B#1} (19#2) #3}
\def\PLBold#1#2#3{Phys. Lett. {\bf#1B} (19#2) #3}
\def\PRD#1#2#3{Phys. Rev. {\bf D#1} (19#2) #3}
\def\PRL#1#2#3{Phys. Rev. Lett. {\bf#1} (19#2) #3}
\def\PRT#1#2#3{Phys. Rep. {\bf#1} (19#2) #3}
\def\ARAA#1#2#3{Ann. Rev. Astron. Astrophys. {\bf#1} (19#2) #3}
\def\ARNP#1#2#3{Ann. Rev. Nucl. Part. Sci. {\bf#1} (19#2) #3}
\def\MPL#1#2#3{Mod. Phys. Lett. {\bf #1} (19#2) #3}
\def\ZPC#1#2#3{Zeit. f\"ur Physik {\bf C#1} (19#2) #3}
\def\APJ#1#2#3{Ap. J. {\bf #1} (19#2) #3}
\def\AP#1#2#3{{Ann. Phys. } {\bf #1} (19#2) #3}
\def\RMP#1#2#3{{Rev. Mod. Phys. } {\bf #1} (19#2) #3}
\def\CMP#1#2#3{{Comm. Math. Phys. } {\bf #1} (19#2) #3}
\relax
%
%
%
\def\beq{\begin{equation}}
\def\eeq{\end{equation}}
\def\bea{\begin{eqnarray}}
\def\eea{\end{eqnarray}}
%
%
%
\newc{\ie}{{\it i.e.}}          \newc{\etal}{{\it et al.}}
\newc{\eg}{{\it e.g.}}          \newc{\etc}{{\it etc.}}
\newc{\cf}{{\it c.f.}}
\def\smuon{{\tilde\mu}}
\def\neut{{\tilde N}}
\def\char{{\tilde C}}
\def\bino{{\tilde B}}
\def\wino{{\tilde W}}
\def\higgsino{{\tilde H}}
\def\sneut{{\tilde\nu}}
%
%
%
%
\def\slash#1{\rlap{$#1$}/} 
\def\Dsl{\,\raise.15ex\hbox{/}\mkern-13.5mu D} 
\def\delsl{\raise.15ex\hbox{/}\kern-.57em\partial}
\def\Ksl{\hbox{/\kern-.6000em\rm K}}
\def\Asl{\hbox{/\kern-.6500em \rm A}}
\def\Qsl{\hbox{/\kern-.6000em\rm Q}}
\def\gradsl{\hbox{/\kern-.6500em$\nabla$}}
%
%
%
\def\bar#1{\overline{#1}}
\def\vev#1{\left\langle #1 \right\rangle}
%

\begin{titlepage} 
\begin{flushright}
OSU-HEP-02-09\\
June 2002\\
\end{flushright}
\vskip 2cm
\begin{center}
{\large\bf Higgs-Mediated $\tau\to3\mu$ in the Supersymmetric Seesaw Model
}
\vskip 1cm
{\normalsize\bf
K.S.~Babu$^{(a)}$ and Christopher Kolda$^{(b)}$ \\
\vskip 0.5cm
$^{(a)}${\it Department of Physics, Oklahoma State University\\
Stillwater, OK~~74078, USA}\\ 
~~ \\
$^{(b)}${\it Department of Physics, University of Notre Dame\\
Notre Dame, IN~~46556, USA}\\[0.1truecm]
}

\end{center}
\vskip .5cm

\begin{abstract}
Recent observations of neutrino oscillations imply non-zero neutrino
masses and flavor violation
in the lepton sector, most economically explained by the seesaw mechanism.
Within the context of supersymmetry, lepton flavor
violation (LFV) among the neutrinos can be communicated by
renormalization group flow to the sleptons and from there to
the charged leptons. We show that LFV can appear in the couplings of
the neutral Higgs bosons, an effect that is strongly enhanced at
large $\tan\beta$. In particular, we calculate the branching fraction for
$\tau\to3\mu$ and $\mu\to 3e$ mediated by Higgs 
and find that they can be as large as $10^{-7}$ and $5\times 10^{-14}$
respectively.
These mode, along with $B^0\to\mu\mu$, can provide important evidence
for supersymmetry before direct discovery of supersymmetric partners occurs.
Along with $\tau\to\mu\gamma$ and $\mu\to e\gamma$, they
can also provide key insights into the form of the neutrino Yukawa
mass matrix.
\end{abstract}

\end{titlepage}

\setcounter{footnote}{0}
\setcounter{page}{1}
\setcounter{section}{0}
\setcounter{subsection}{0}
\setcounter{subsubsection}{0}


Over the last several years, evidence from a number of experiments,
notably SuperKamiokande
and SNO, has pointed conclusively to the existence of
neutrino oscillations in atmospheric~\cite{atmospheric} and 
solar neutrinos~\cite{solar}
and, by implication, to non-zero neutrino masses.
Within the context of the Standard Model (SM), the most attractive
explanation for the observed neutrino masses is the ``seesaw'' 
mechanism~\cite{seesaw}:
right-handed neutrinos are introduced in order to couple with
left-handed neutrinos through SU(2)$\times$U(1)-violating 
Dirac mass terms, $m_D$,
while also receiving large, SU(2)$\times$U(1)-invariant 
Majorana masses, $M_R$. The resulting spectrum consists of
heavy neutrinos with masses $\sim M_R$
which are primarily right-handed, and neutrinos with extremely small masses
$m_\nu\sim m_D^2/M_R$ which are primarily left-handed. 
The seesaw mechanism also has the virtue that it fits elegantly
inside a grand unified theory (GUT), such as SO(10).

Within SO(10), the Dirac neutrino masses are predicted to be of order
the corresponding up-quark masses; for example, $(m_D)_{\nu_\tau}$ would
be roughly $100$ to $200\gev$. 
Atmospheric neutrino data favors a $\nu_\tau$ mass of about
$0.04\ev$~\cite{nufits}. Thus one
finds a right-handed Majorana mass, $M_R$, of order $10^{14}\gev$, several
orders of magnitude below the Planck scale ($\mpl$).

Majorana neutrino masses and neutrino oscillations imply lepton flavor 
violation (LFV). But within the SM, flavor violation in charged lepton
processes is necessarily generated by irrelevant operators and is
therefore suppressed by powers of $1/M_R$ (or equivalently powers 
of $m_\nu$).
Within supersymmetric extensions of the SM, however, this is no
longer true. In the minimal supersymmetric standard model (MSSM)
(which we will henceforth take to be augmented 
with heavy right-handed neutrinos, 
$\nu_R$)
LFV can be communicated directly from $\nu_R$
to the sleptons by relevant operators and from there 
to the charged leptons. LFV is then suppressed by powers
of $1/M_{\rm SUSY}$ instead of $1/M_R$, with $M_{\rm SUSY}\ll M_R$.
The initial communication is done most
economically through renormalization group flow of the slepton mass
matrices at energies between $\mpl$ and $M_R$. 
Though the scale $M_R$ is far above
the weak scale, the presence of the $\nu_R$ at scales above $M_R$
leaves a
(non-decoupling) imprint on the mass matrices of the sleptons which is
preserved down to the weak scale. This effect has been used 
to predict
large branching fractions for $\tau\to\mu\gamma$ and $\mu\to e\gamma$
within the MSSM~\cite{gammaold,moroi,gamma}.

In this letter we demonstrate a new way in which the imprint of
LFV on the slepton mass matrices can be communicated to
charged leptons through the exchange of Higgs bosons,
providing the possibility of new and observable
flavor violation in the leptonic sector. We will
demonstrate that the decay $\tau\to 3\mu$ is a particularly sensitive
probe of LFV at large $\tan\beta$ (the ratio  
$\langle H_u\rangle/\langle H_d\rangle$), with a branching fraction
scaling as $\tan^6\beta$. The process $\mu\to 3e$ can also proceed via
Higgs exchange with the same $\tan^6\beta$-dependence; though the
branching fraction is suppressed by the small electron Yukawa
coupling, it is still within the grasp of experiment. We will
concentrate our discussion on the $\tau$ decay and come back to the
$\mu$ decay in the final discussion.

At the end we will also briefly discuss the related
processes $\tau\to 
\mu\mu e$ and $\tau\to\mu ee$.
We find no new enhancement for the decay $\tau\to 3e$.
These decay modes are (like $\tau\to\mu\gamma$)
only logarithmically sensitive to $M_R$, but (unlike
$\tau\to\mu\gamma$) do not decouple in the large $m_\slepton$ mass
limit. Rather, these modes decouple in the limit that the pseudoscalar
Higgs boson becomes heavy, $m_A\to\infty$, thus providing
complementary information on the supersymmetric (SUSY) spectrum.

~

{\bf Flavor Violation among the Sleptons.}
In the leptonic sector, we begin with a Lagrangian:
\beq
-{\cal L}=\bar E_R Y_E L_L H_d + \bar\nu_R Y_\nu L_L
+{\textstyle\frac12} \nu_R^\top M_R\, \nu_R
\eeq
where $E_R$, $L_L$ and $\nu_R$ represent $3\times 1$
matrices in flavor space 
of right-handed charged leptons, left-handed lepton doublets
and right-handed neutrinos, and $Y_E$, $Y_\nu$ and $M_R$ are $3\times 3$
matrices in flavor space; for example, $E_R=(e_R, \mu_R, \tau_R)^\top$.
This Lagrangian clearly violates both family and total lepton number due
to the presence of the $\nu_R$ Majorana mass term.
We can choose to work in a basis in which both $Y_E$ and $M_R$ have
been diagonalized, but
$Y_\nu$ remains an arbitrary, complex matrix.

Within the SM, $O(1)$ flavor violation in the neutrinos does not translate
into appreciable flavor violation 
in the charged lepton sector due to $1/M_R$ suppressions. But this is
not true in the slepton sector. The SUSY-breaking slepton masses are
unprotected by chiral symmetries and are therefore sensitive to
physics at all mass scales between $m_{\Slepton}$ 
and the scale, $M$, at which
SUSY-breaking is communicated to the visible sector, assuming $M>M_R$.
This can be seen
by examining the renormalization group equation for $m_{\Slepton}^2$
at scales above $M_R$:
\bea
\frac{d}{d\log Q}(m^2_\Slepton)_{ij} &=& \left(
\frac{d}{d\log Q}(m^2_\Slepton)_{ij}\right)_{\rm MSSM} \\
&&+\frac{1}{16\pi^2}\left[m^2_\Slepton Y_\nu^\dagger Y_\nu
+Y_\nu^\dagger Y_\nu m^2_\Slepton
+2(Y_\nu^\dagger m^2_{\sneutrino_R} Y_\nu + m^2_{H_u}Y_\nu^\dagger Y_\nu
+A_\nu^\dagger A_\nu)\right]_{ij} \nonumber
\eea
where the first term represents the ($L$-conserving) 
terms present in the usual MSSM at
scales below $M_R$. Because $Y_\nu$ is off-diagonal, it will generate
flavor-mixing in the slepton mass matrix.
We can solve this equation approximately for the flavor-mixing piece:
\beq
\left(\Delta m^2_\Slepton\right)_{ij} \simeq
-\frac{\log(M/M_R)}{16\pi^2}\left(6m_0^2(Y_\nu^\dagger Y_\nu)_{ij}
+2\left(A_\nu^\dagger A_\nu\right)_{ij}\right)
\eeq
where $m_0$
is a common scalar mass evaluated at the scale $Q=M$, and $i\neq j$. 
If we further
assume that the $A$-terms are proportional to Yukawa matrices, then:
\beq
\left(\Delta m^2_\Slepton\right)_{ij} \simeq \xi \left(Y_\nu^\dagger
Y_\nu\right)_{ij}
\eeq
where
\beq
\xi =
-\frac{\log(M/M_R)}{16\pi^2}(6+2a^2)m_0^2.
\label{log}
\eeq
and $a$ is $O(1)$.
In the simplest SUSY-breaking scenarios, gravity plays the
role of messenger and $M=\mpl$, so that the logarithm in
Eq.~(\ref{log}) is roughly 10.

What does experiment tell us about the values of these matrices?
Global fits to neutrino data favor large mixing between
the $\nu_\mu$ and $\nu_\tau$, and also between $\nu_e$ and
$\nu_\mu$~\cite{nufits}. 
We will consider the following form for $m_\nu$ which
provides an excellent fit to existing neutrino data and can be motivated
by theory~\cite{altarelli}:
\beq
m_\nu \propto \left(\begin{array}{ccc} \eps & \eps & \eps \\
\eps & 1 & 1 \\ \eps & 1 & 1 \end{array} \right)
\label{mnu}
\eeq
where $\eps$ is a small parameter $\sim0.1$.
If we further assume that $M_R$ is an identity matrix, then
$Y_\nu^\dagger Y_\nu$ will also have the form of Eq.~(\ref{mnu}).
Another interesting possibility is provided by
GUT models with lopsided mass matrices for charged
leptons~\cite{lopsided}; such models have 
$(Y_E)_{32}\simeq (Y_E)_{33}$ and 
lead to a light neutrino mass matrix as in Eq.~(\ref{mnu}) with
$(Y_\nu)_{32}\simeq (Y_\nu)_{33}\simeq y_t$, where $y_t$ is the top 
Yukawa coupling. In either case, 
the $Y_\nu^\dagger Y_\nu$ 
has $O(1)$ flavor violation in the $\nu_\tau$--$\nu_\mu$ 
sector if $M_R\simeq 10^{14}\gev$.

~

{\bf Higgs-Mediated Flavor Violation.} 
Unlike the SM, the MSSM is not protected against the possibility of
flavor-changing neutral currents (FCNCs) 
mediated by neutral Higgs bosons. Though the MSSM is a type-II
two-Higgs doublet model at tree level, this structure is not protected
by any symmetry. In particular, the presence of a non-zero $\mu$-term,
coupled with SUSY-breaking, is enough to induce non-holomorphic Yukawa
interactions for the quarks and leptons. In the quark sector this was
discovered by Hall, Rattazzi and Sarid~\cite{hrs}; terms of the
form $\bar Q_Lu_RH_d^\dagger$ and $\bar Q_Ld_R H_u^\dagger$ were found,
the latter providing a significant correction to the
$b$-quark mass at large $\tan\beta$. We have shown in a previous
letter~\cite{prl} that these terms allow the neutral Higgs bosons to
mediate FCNCs, in particular $B\to\mu\mu$. There we argued that
branching fractions predicted at large $\tan\beta$ can be probed at
Run~II of the Tevatron. (For recent analyses of Higgs-mediated
$B\to\mu\mu$, see Ref.~\cite{bmuanalysis}.)

The two leading diagrams considered in Refs.~\cite{hrs,prl} as a
source for non-holomorphic quark couplings are not present in the
leptonic sector since they involve gluinos and top squarks inside the
loops. 
However there are additional diagrams which are present in
the leptonic sector~\cite{cpv} involving loops of sleptons and
charginos/neutralinos; a subset of these is shown in 
Fig.~1. Given a source of non-holomorphic
couplings and LFV among the sleptons, 
Higgs-mediated LFV is unavoidable.

For those familiar with the quark sector FCNC calculation of
Ref.~\cite{prl}, we note that 
the present calculation is not very different. In particular,
the contributions considered here will be similar in most ways to
those in Ref.~\cite{prl} which were generated by a squark mixing
insertion in the gluino-sbottom loop diagram.

We begin by writing the effective Lagrangian for the couplings of the
charged leptons to the neutral Higgs fields:
\beq
-{\cal L}=\bar E_R Y_E E_{L} H_d^0 +\bar E_R Y_E \left(\eps_1{\bf 1}
+\eps_2 Y_\nu^\dagger Y_\nu\right) E_{L} H_u^{0*} + h.c.
\label{FCLag}
\eeq
The first term is the usual Yukawa coupling, while the second term
arises from the non-holomorphic loop corrections. LFV
results
from our inability to simultaneously diagonalize $Y_E$ and the
$\eps_2 Y_EY_\nu^\dagger Y_\nu$ term\footnote{There are additional
terms which can be
written, but these are either $L$-conserving or subleading in the
LFV calculation that follows.}, just as in 
Ref.~\cite{prl}; as $\eps_2\to0$,
LFV in the Higgs sector will disappear.

The diagrams which contribute to $\eps_2$ are shown in Fig.~1.
Each diagram contains a single insertion of $\Delta m^2_\Slepton$
which introduces LFV into the process. Without this
insertion, these diagrams would have
a trivial flavor structure and would not
contribute to $\eps_2$ or to LFV.
But the $\Delta m^2_\Slepton$ insertion introduces a $Y_\nu^\dagger
Y_\nu$ into the diagram, yielding a contribution to $\eps_2$. We can
approximate the contributions of the diagrams in Fig.~1 by inserting a
single $\Delta m_{\Slepton}^2$ mass insertion onto each of the
internal $\tilde E_L$ lines. We will also treat the higgsinos and
gauginos as approximate mass eigenstates.
\begin{figure}
\centering
\epsfysize=2.2truein
\hspace*{0in}
\epsffile{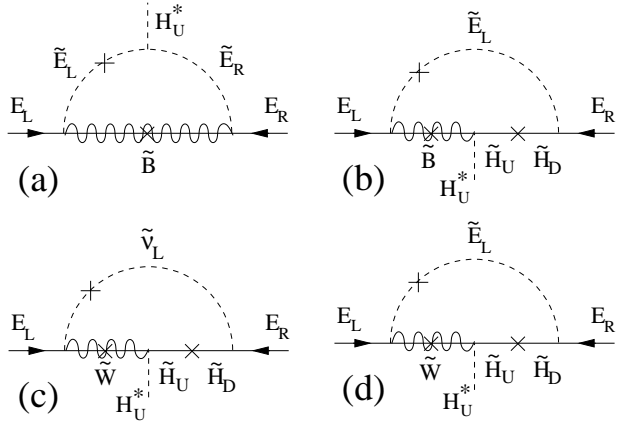}
\caption{Diagrams that contribute to $\eps_2$. The crosses on the
internal slepton lines represent LFV mass insertions due to
loops of $\nu_R$.
}
\label{fig1}
\end{figure}
For
diagram~1(a), the contribution to $\eps_2$ is 
\beq
\eps_{2a} \simeq 
\frac{\alpha'}{4\pi}\xi\mu M_1 f_2\left(M_1^2,m_{\slepton_L}^2,
m_{\stau_L}^2,m_{\slepton_R}^2\right)
\label{eq:2a} 
\eeq 
where $\slepton=\smuon$ or $\selectron$.
Diagram~1(b) provides a contribution given by
\beq
\eps_{2b}\simeq-\frac{\alpha'}{8\pi}\xi\mu M_1 f_2
\left(\mu^2, m^2_{\slepton_L}, m^2_{\stau_L},M_1^2\right).
\label{eq:2b}
\eeq
Diagram~1(c) yields
\beq
\eps_{2c} \simeq  \frac{\alpha_2}{4\pi}\xi\mu M_2
f_2\left(\mu^2,m_{\sneutrino_\ell}^2,m_{\sneutrino_\tau}^2,M_2^2\right).
\label{eq:2c}
\eeq
Finally, the contribution of diagram~1(d) is found to be
\beq
\eps_{2d} \simeq  \frac{\alpha_2}{8\pi}\xi\mu M_2 f_2\left(\mu^2,
m^2_{\slepton_L}, m^2_{\stau_L}, M_2^2\right). 
\label{eq:2d}
\eeq
In these equations, $M_{1,2}$ are the U(1) and SU(2) gaugino masses,
$\xi$ is defined in Eq.~(\ref{log}),
and the function $f_2$ is defined such that
{\small
\beq
-f_2(a,b,c,d) \equiv
\frac{a\log(a)}{(a-b)(a-c)(a-d)} +\frac{b\log(b)}{(b-a)(b-c)(b-d)}
+(a\leftrightarrow c,
b\leftrightarrow d) .
\eeq
}

\noindent
The function $f_2$ is positive definite and we note several
interesting limits in its behavior. When
$a=b=c=d$, $f_2(a,a,a,a)=1/(6a^2)$; and when $a\gg b=c=d$, then
$f_2(a,b,b,b) = 1/(2ab)$. The total contribution to
$\eps_2$ is the sum of the individual pieces listed above.
Whether we identify the slepton $\tilde\ell=\tilde\mu$ or 
$\tilde e$ depends on the decay
process we are considering.

It is clear from the Lagrangian in Eq.~(\ref{FCLag}) that the charged lepton
masses cannot be diagonalized in the same basis as their Higgs couplings.
This will allow neutral Higgs bosons to mediate LFV
processes with rates proportional to $\eps_2^2$. But in order to 
proceed, we will choose a specific process to study, namely $\tau\to
3\mu$. Our discussions will generalize to other related
processes (such as $\tau\to \mu ee$) very easily.

We will only mention in passing the contributions to $\eps_1$, since
they do not induce LFV. The diagrams which contribute to
$\eps_1$ are (mostly) those of Fig.~1 without the slepton mass
insertion.  The contributions of
these diagrams to $\eps_1$ are found to be~\cite{cpv}:
\bea
\eps_{1}&=&\frac{\alpha'}{8\pi}\mu M_1 \left[2f_1\left(M_1^2,m_{\slepton_L}^2,
m_{\slepton_R}^2\right) -
f_1\left(M_1^2,\mu^2,m^2_{\slepton_L}\right) + 2f_1\left(M_1^2,\mu^2,
m^2_{\slepton_R}\right)\right] \nonumber \\ 
&&+\frac{\alpha_2}{8\pi}\mu M_2\left[f_1\left(\mu^2,
m^2_{\slepton_L}, M_2^2\right) 
+ 2f_1\left(\mu^2,m_{\sneutrino}^2,M_2^2\right)\right]
\label{eq:eps1} 
\eea
where
\beq
-f_1(a,b,c)\equiv \frac{ab\log(a/b)+bc\log(b/c)+ca\log(c/a)}
{(a-b)(b-c)(c-a)}.
\eeq
These terms will generate a mass shift for the charged leptons that
will appear in our final formulae as a second-order effect.

~

{\bf Flavor--Violating Tau Decays.}
We begin by extracting the $\bar\tau_R\,\mu_L$ terms in the effective
Lagrangian of Eq.~(\ref{FCLag}). The algebra is similar to that in
Ref.~\cite{prl} and so we skip directly to the result. The relevant
LFV interaction has the form:
\beq
-{\cal L}\simeq (2G_F^2)^{1/4}
\frac{m_\tau \kappa_{32}}{\cos^2\beta}
\left(\bar\tau_R\,\mu_L\right)
\left[\cos(\beta-\alpha)
h^0 - \sin(\beta-\alpha) H^0 - iA^0\right]+h.c.
\label{finalL}
\eeq
where 
\beq
\kappa_{ij} = -\frac{\eps_2}{
\left[1+(\eps_1+\eps_2
(Y_\nu^\dagger Y_\nu)_{33})\tan\beta\right]^2 }
\left(Y_\nu^\dagger Y_\nu\right)_{ij}.
\eeq
(The Lagrangian for $(\bar\tau_R\, e_L)$--Higgs can be derived from this
by replacing $\kappa_{32}$ with $\kappa_{31}$.) 

Given an LFV $(\bar\tau_R\, e_L)$--Higgs interaction, 
the decay $\tau\to3\mu$ can be generated
via exchange of $h^0$, $H^0$ and $A^0$ as in Fig.~2. 
\begin{figure}
\centering
\epsfysize=0.8truein
\hspace*{0in}
\epsffile{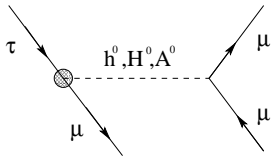}
\caption{The Feynman diagram contributing to $\tau\to3\mu$. The shaded
interaction vertex is the new vertex derived in Eq.~\ref{finalL}.
}
\label{fig2}
\end{figure}
The diagram in Fig.~2 is straightforward to calculate. We derive a
branching fraction for $\tau\to3\mu$ of 
\beq
\mbox{Br}(\tau\to3\mu)= 
\frac{G_F^2 m_\mu^2 m_\tau^7 \tau_\tau}{768\pi^3 m_A^4}
 \kappa_{32}^2 \tan^6\beta
\eeq
where $\tau_\tau$ is the $\tau$ lifetime. To derive this formula, we
took the large $m_A$ limit in which $\alpha\to\beta-\pi/2$.

How large can this branching fraction be? Consider the case in which 
$\mu= M_1= M_2= m_{\slepton}=
m_{\sneutrino}$, $M_R=10^{14}\gev$ and $(Y_\nu^\dagger Y_\nu)_{32}=1$.
Then $\eps_2 \simeq 4\times 10^{-4}$ and
\beq
\mbox{Br}(\tau\to3\mu) \simeq (1\times 10^{-7})\times\left(\frac{\tan\beta}
{60}\right)^6\times\left(\frac{100\gev}{m_A}\right)^4
\eeq
which puts it into the regime that is experimentally accessible
at B-factories over the next few years. At LHC and SuperKEKB,
limits in the region of $10^{-9}$ should be achievable~\cite{exp},
allowing a deeper probe into the parameter space.
We can also do better if 
$\mu\gg M_{1,2}\simeq m^2_{\slepton}$. Then the bino contribution is
enhanced by a factor $\mu/M_1$; for $M_1\simeq 100\gev$ and $\mu\simeq
1\tev$, one can get $\eps_2\simeq 8\times 10^{-4}$, resulting in a
branching fraction 4 times that stated above. However we note that the
value of $\eps_2\simeq 4\times 10^{-4}$ is remarkably stable to
changes in the SUSY spectrum apart from this large-$\mu$ option.

~

{\bf Discussion.}
We have demonstrated that LFV in the sleptons can generate large LFV
in the couplings of leptons to neutral Higgs bosons. It is already
well-known that sleptonic flavor violation can induce LFV in certain
magnetic moment transitions such as $\tau\to\mu\gamma$, so it 
is useful to compare this to the $\tau\to 3\mu$ decay which we just
found. We first note that the two
decays possess very different decoupling behavior so that either
one could be
large while the other is too small to observe. The effective operator
for $\tau\to 3\mu$ is dimension-6:
$(1/m_A^2)\bar\tau\mu\bar\mu\mu$
The $\tau\to\mu\gamma$ operator is formally dimension-5, but chiral
symmetry requires an $m_\tau$ insertion, so that the operator is
actually dimension-6:
$(m_\tau/M_{\rm SUSY}^2)\bar\tau\sigma^{\mu\nu}\mu F_{\mu\nu}$
where $M_{\rm SUSY}$ represents the heaviest mass scale to enter the 
slepton-gaugino loops. If sleptons and gauginos are light and $A^0$ is
heavy, then $\tau\to\mu\gamma$ would tend to dominate; in the opposite
limit and with $\tan\beta$ large, $\tau\to 3\mu$ would
dominate. Because of this different decoupling behavior, it is
impossible to correlate the two decays without choosing a specific model.
Turning this around, observation of one or both of these decays can
provide insight into the fundamental SUSY-breaking parameters.

The presence of the $\tau\to\mu\gamma$ operator can also 
lead to $\tau\to 3\mu$ if the photon goes off-shell. 
However, for this operator the relation between the two
branching fractions is roughly model-independent~\cite{moroi}:
\beq
\frac{\mbox{Br}(\tau\to3\mu)}{\mbox{Br}(\tau\to\mu\gamma)}
\simeq 0.003\quad\mbox{(no Higgs-mediation)}.
\eeq
If a ratio much larger than $0.003$ is discovered, then this would be
clear evidence that some new process is generating the $\tau\to3\mu$
decay, with Higgs-mediation a leading contender.

Our calculation so far has also relied on Yukawa matrix
{\it ans\"atze}\/ which reproduce the mass matrix of Eq.~(\ref{mnu}).
But what if we had made a different choice for $Y_\nu^\dagger
Y_\nu$? Another popular option would be the inverted mass hierarchy in
which the (1,2) and (1,3) elements of $Y_\nu^\dagger Y_\nu$ would be
$O(1)$ and the remainder $O(\eps)$~\cite{altarelli}. 
Such a matrix could lead to observable $\tau\to e\mu\mu$.
The constraints coming from $\mu\to
e\gamma$ are already strong; for $M_{\rm SUSY}\simeq 100\gev$ one
finds that $(Y_\nu^\dagger Y_\nu)_{21}$ has
to be $\lsim 10^{-2}$~\cite{ellis}. But in the large $M_{\rm SUSY}$,
large $\tan\beta$ limit, this bound is weakened and $\tau\to e\mu\mu$ 
could dominate.

There is another interesting test of the inverted hierarchy models
presented by Higgs mediation. The decay $\mu\to 3e$ can also proceed
by neutral Higgs exchange. Though the electron Yukawa coupling is
tiny, this is offset by the extreme precision of rare $\mu$-decay
searches. In particular, we find that for $(Y_\nu^\dagger
Y_\nu)_{21}= 1$,
\beq
\mbox{Br}(\mu\to 3e) \simeq (5\times 10^{-14})\times\left(\frac{\tan\beta}
{60}\right)^6\times\left(\frac{100\gev}{m_A}\right)^4.
\eeq
Again, this process can generated by the similar $\mu\to e\gamma$
operator taken off-shell, but there the ratio is again
model-independent:
\beq
\frac{\mbox{Br}(\mu\to 3e)}{\mbox{Br}(\mu\to e\gamma)}
\simeq 0.006\quad\mbox{(no Higgs-mediation)}.
\eeq
Any deviation from this fixed ratio would, as for the taus, be a
strong indication of new physics such as that found here.

Finally, we mention here
that one could also calculate the rate for processes involving
LFV Higgs couplings at both vertices, 
though we will leave this computation to a future work.
For example, a second
way to generate $\tau\to e\mu\mu$ would be to use the
$\bar\tau_R\,\mu_L$--Higgs coupling that we
have been considering in this paper, along with a $\bar\mu_R
e_L$--Higgs coupling. We can also generate $\tau\to ee\mu$ with a
$\bar\tau_R\, e_L$--Higgs coupling at one end and a $\bar\mu_R
e_L$--Higgs coupling at the other. 
But as above, both processes are constrained by non-observation of
$\mu\to e\gamma$ since they require large $(\Delta
m^2_{\Slepton})_{21}$. It is notable that these processes have a
remarkable $\tan^8\beta$ dependence; however the additional powers
are mitigated by additional loop suppressions.

~

\noindent{\bf Acknowledgments.}
We would like to thank the scientific staffs of CERN and DESY for their
hospitality while this work was completed, and especially P.~Zerwas
for organizing a thought-provoking SUSY'02 Conference. CK would also
like to thank D.~Chung and L.~Everett for hosting him at their Physics
Institute-on-Wheels where much of this letter was composed.
This research was supported
by the National Science Foundation under grant PHY00-98791, by the
U.S.~Department of Energy under grants DE-FG03-98ER41076 and 
DE-FG02-01ER45684, and by a grant from the Research Corporation.

\end{document}